\documentclass[reprint,a4paper,english,showpacs,prd]{revtex4-1}   

\usepackage{babel,amsmath,amssymb,dcolumn}
\usepackage[dvips]{graphics}

\begin{document}

\title{Quasinormal behavior of massless scalar field perturbation \\ in
  Reissner-Nordstr\"{o}m anti-de Sitter spacetimes} 

\author{Bin Wang}
\email{binwang@fudan.ac.cn}
\affiliation{Department of Physics, Fudan University, 200433 Shanghai}

\author{Chi-Yong Lin}
\email{lcyong@mail.ndhu.edu.tw}
\affiliation{Department of Physics, National Dong Hwa University,
  Shoufeng, 974 Hualien} 

\author{C. Molina}
\email{cmolina@usp.br}
\affiliation{Instituto de F\'{\i}sica, Universidade de S\~{a}o Paulo,
  C.P.66.318, CEP 05315-970,  S\~{a}o Paulo-SP, Brazil}

\begin{abstract}
We present a comprehensive study of the massless scalar field
perturbation in the Reissner-Nordstr\"{o}m anti-de Sitter (RNAdS)
spacetime and compute its quasinormal modes (QNM). For the lowest
lying mode, we confirm and extend the dependence of the  QNM
frequencies on the black hole charge got in previous works. In near
extreme limit, under the scalar perturbation we find that the
imaginary part of the frequency tends to zero, which is consistent
with the previous conjecture based on electromagnetic and axial
gravitational perturbations. For the extreme value of the charge, the
asymptotic field decay is dominated by a power-law tail, which shows
that the extreme black hole can still be stable to scalar
perturbations. We also study the higher overtones for the RNAdS  black
hole and find large variations of QNM frequencies with the overtone
number and black hole charge. The nontrivial dependence of frequencies
on the angular index $\ell$ is also discussed.    
\end{abstract}


\maketitle

%%%%%%%%%%%%%%%%%%%%%%%%%%%%%%%%%%%%%%%%%%%%%%%%%%%%%%%%%%%%%%%%%%%%%%%%%%%%%%%
\section{Introduction} 
%%%%%%%%%%%%%%%%%%%%%%%%%%%%%%%%%%%%%%%%%%%%%%%%%%%%%%%%%%%%%%%%%%%%%%%%%%%%%%%

It is well known that under perturbations  the surrounding geometry of
a black hole will experience damped oscillations with frequencies and
damping times entirely determined by the black hole parameters. These
oscillations are called  ``quasinormal modes'' (QNM). They carry unique
fingerprints of black holes which would lead to the direct
identification of the black hole existence. Due to the potential
astrophysical interest, extensive study of QNM of black holes in
asymptotically flat spacetimes have been performed for the last few
decades (for comprehensive reviews see \cite{Nollert-99,Kokkotas-99} and
references therein). Considering the case when the black hole is
immersed in an expanding universe, the QNMs of black holes in de
Sitter space have also been investigated \cite{deSitter_1,deSitter_2}.  

Motivated by the recent discovery of the anti-de Sitter/conformal
field theory (AdS/CFT) correspondence, the investigation of QNMs in
anti-de Sitter (AdS) spacetimes became appealing in the past years. It
was argued that the QNMs of AdS black holes have direct interpretation
in terms of the dual conformal field theory. The first study of the
QNMs in AdS spaces was performed by Chan and Mann
\cite{Chan_Mann}. 
Subsequently, Horowitz and Hubeny suggested a
numerical method to calculate the QN frequencies directly and made a
systematic investigation of QNMs for scalar perturbation on the 
background of Schwarzschild AdS (SAdS) black holes \cite{Horowitz-00}.
Considering that the RNAdS solution provides a better framework than
the SAdS geometry and may contribute significantly to our
understanding of space and time, the Horowitz-Hubeny numerical method
was generalized to the study of QNMs of RNAdS black holes in \cite{Wang-00}
and later crosschecked by using the time evolution approach
\cite{Wang-01}. In addition to the scalar perturbation, gravitational and
electromagnetic perturbations in AdS black holes have also attracted
attention \cite{Cardoso,Berti-03}. Other works on QNMs in AdS
spacetimes can be found in 
\cite{Konoplya-02, Birmingham,Zhu_Wang_Mann_Abdalla,
  AdS_SeveralAuthors,Cardoso_Lemos,Cardoso_Konoplya_Lemos}.    

Recently in \cite{Berti-03} Berti and Kokkotas used the frequency-domain
method and restudied the scalar perturbation in RNAdS black
holes. They verified most of our previous numerical results in
\cite{Wang-00,Wang-01}, however, disapproved the property of the real part of
the QN frequency with the increase of the black hole charge. Our first
aim in this paper is to check their results by using both improved
Horowitz-Hubeny method and time evolution approach. 

As was pointed out in \cite{Wang-00} and later supported in
\cite{Berti-03}, the Horowitz-Hubeny method breaks down for large
values of the charge. To study the QNMs in the near extreme and
extreme RNAdS backgrounds, we need to count on time evolution
approach. Employing an improved numerical method, we will show that
the problem with minor instabilities in the form of ``plateaus'', which
were observed in \cite{Wang-01}, is overcome. We obtain the precise
QNMs behavior in the highly charged RNAdS black holes.   

\newpage

In addition to the study of the lowest lying QNMs, it is interesting
to study the higher overtone QN frequencies for scalar
perturbations. The first attempt was carried out in
\cite{Cardoso_Konoplya_Lemos}. In the present work we would
like to extend the study to the RNAdS backgrounds.  

It was argued that the dependence of the QN frequencies on the angular
index $\ell$ is extremely weak \cite{Berti-03}. This was also claimed in
\cite{Cardoso_Konoplya_Lemos}. Using our numerical results we
will show that this weak dependence on the angular index is not
trivial.  

The plan of the paper is as follows. In Sec.II we review the
background metric, display the scalar wave equations and briefly
present our numerical methods. In Sec.III we describe in detail our
numerical results. The conclusions and discussions will be presented
in Sec.IV.

%%%%%%%%%%%%%%%%%%%%%%%%%%%%%%%%%%%%%%%%%%%%%%%%%%%%%%%%%%%%%%%%%%%%%%%%%%%%%%%
\section{Equations and numerical methods} 
%%%%%%%%%%%%%%%%%%%%%%%%%%%%%%%%%%%%%%%%%%%%%%%%%%%%%%%%%%%%%%%%%%%%%%%%%%%%%%%

%%%%%%%%%%%%%%%%%%%%%%%%%%%%%%%%%%%%%%%%%%%%%%%%%%%%%%%%%%%%%%%%%%%%%%%%%%%%%%%
\subsection{Geometry and fields}
%%%%%%%%%%%%%%%%%%%%%%%%%%%%%%%%%%%%%%%%%%%%%%%%%%%%%%%%%%%%%%%%%%%%%%%%%%%%%%%

The metric describing a charged, asymptotically anti-de Sitter spherical black
hole, written in spherical coordinates, is given by
\begin{equation}
ds^{2} = -h(r) dt^{2} + h(r)^{-1} dr^{2} +
r^{2}(d\theta^{2}+\sin^{2}\theta d\phi^{2}) \ ,
\label{metric}
\end{equation}
where the function $h(r)$ is
\begin{equation}
h(r) = 1 - \frac{2m}{r} + \frac{Q^{2}}{r^{2}} - \frac{\Lambda r^{2}}{3} \ .
\end{equation}
We are assuming a negative cosmological
constant, usually written as $\Lambda = -3/R^{2}$. The integration
constants $m$ and $Q$ are the black hole mass and electric charge
respectively. The extreme value of the black hole charge, $Q_{max}$,
is given by the function of the event horizon radius in the form
$Q_{max}^2=r_+^2(1+3r_+^2/R^2)$. 

The spacetime causal structure depends on the zeros of $h(r)$.
Changing the parameters $m$, $Q$ and $R$, the function $h(r)$ may
have none, one, or two positive zeros. In the  Reissner-Nordstr\"om-anti-de
Sitter case, $h(r)$ has two simple real, positive roots  ($r_{+}$ and $%
r_{-}$) and two complex roots ($r_{1}$ and $r_{1}^{*}$, asterisk denoting
complex conjugation). In the
so-called extreme Reissner-Nordstr\"om-anti-de Sitter case, $h(r)$ has
one double positive zero (also denoted $r_{+}$) and two complex roots
($r_{1}^{ext}$ and $\left(r_{1}^{ext}\right)^{*}$). The horizons
$r_{-}$ and $r_{+}$ with $r_{-}<r_{+}$, are called Cauchy and event
 horizons, respectively. In this work we will treat the dynamics of
fields in the black hole exterior, the submanifold given by the patch
$T_{+}=\left\{ (t,r,\theta,\phi),r > r_{+} \right\}$.  

In the region $T_{+}$, we define a ``tortoise coordinate'' $r^*(r)$ in the
usual way,
\begin{equation}
r^*(r) = \int\frac{dr}{h(r)} \,\, .
\end{equation} 
Assuming $h(r)$ has two distinct positive roots \linebreak (nonextreme case),
the tortoise coordinate is given by
\begin{multline}
r^*(r) = \frac{1}{2\kappa_{+}} \ln \left(r-r_{+}\right) -
 \frac{1}{2\kappa_{-}} \ln\left(r-r_{-}\right) \\
 - A \ln(r^{2}+pr+q) 
  + \frac{2(B + Ap)}{\sqrt{4q-p^{2}}} \\
 \times \left[\arctan\left(\frac{2r+p}{\sqrt{4q-p^{2}}}\right) -
 \frac{\pi}{2}\right] \,\, ,
\label{ne_tortoise}
\end{multline}
with the constants $\kappa_{+}$, $\kappa_{-}$, $A$, $B$, $p$
and $q$ given by 
\begin{equation}
\frac{1}{2\kappa_{+}} = \frac{R^{2}r_{+}^{2}}{(r_{+} -
  r_{-})(3r_{+}^{2} + r_{-}^{2} + 2r_{+}r_{-}+R^{2})} \,\, ,
\end{equation}
\begin{equation}
\frac{1}{2\kappa_{-}} = \frac{R^{2}r_{-}^{2}}{(r_{+} -
  r_{-})(3r_{-}^{2} + r_{+}^{2} + 2r_{+}r_{-} + R^{2})} \,\, ,
\end{equation}
\begin{equation}
A = \frac{R^{2} (r_{+} + r_{-})(r_{+}^{2} +
  r_{-}^{2} + 2r_{+}r_{-} + R^{2})}{2(3r_{+}^{2} + r_{-}^{2} +
  2r_{+}r_{-} + R^{2})(3r_{-}^{2}+r_{+}^{2}+2r_{+}r_{-}+R^{2})}
  \,\, ,
\end{equation}
\begin{equation}
B = \frac{R^{2}(r_{+}^{2} + r_{-}^{2} + R^{2})(r_{+}^{2} + r_{-}^{2} +
  r_{+}r_{-} + R^{2})}{(3r_{+}^{2} + r_{-}^{2} + 2r_{+}r_{-} +
  R^{2})(3r_{-}^{2}+r_{+}^{2} + 2r_{+}r_{-} + R^{2})} \,\, ,
\end{equation}
\begin{equation}
p = r_{+} + r_{-} \,\, , \,\, q = R^{2} + r_{+}^{2} + r_{-}^{2} +
r_{+} \,\, . 
\end{equation}
The constant of integration in the expression (\ref{ne_tortoise}) was
chosen so that $\lim_{r\rightarrow\infty} r^*(r) = 0$. 

Consider now a scalar perturbation field $\Phi$ obeying the massless
Klein-Gordon equation  
\begin{equation}
\Box\Phi = 0 \ . 
\label{boxphiequalzero}
\end{equation}
The usual separation of variables in terms of a radial field and a
spherical harmonic $\textrm{Y}_{\ell,m}(\theta,\varphi)$,
\begin{equation}
\Phi=\sum_{\ell\,m} \frac{1}{r}
\Psi(t,r)\textrm{Y}_{\ell m}(\theta,\phi)  \ , 
\label{Ansazs_field}
\end{equation}
leads to Schr\"{o}dinger-type equations in the tortoise coordinate
for each value of $\ell$. Introducing the null coordinates $u = t -
r^*$ and $v =  t + r^*$, the field equation is given by 
\begin{equation}
- 4 \frac{\partial^{2}\Psi}{\partial u \partial v}
 = V(r) \Psi \ ,
\label{u_v_wave_equation}
\end{equation}
where the effective potential $V$ is 
\begin{equation}
V(r)=h(r)\left[\frac{\ell(\ell + 1)}{r^2}+\frac{2m}{r^{3}} -
  \frac{2Q^2}{r^{4}} + \frac{2}{R^{2}}\right] \ . 
\end{equation}
Wave equation (\ref{u_v_wave_equation}) is useful to study the time
evolution of the scalar perturbation, in the context of an initial
characteristic value problem, explored in this work.

In terms of the ingoing Eddington coordinates $(v,r)$ and separating
the scalar field in a product form as   
\begin{equation}
\Phi= \sum_{\ell m} \frac{1}{r}
\psi(r) \textrm{Y}_{\ell m}(\theta,\phi) e^{-i\omega v},  
\end{equation}
the minimally-coupled scalar wave equation (\ref{boxphiequalzero}) may
thereby be reduced to  
\begin{equation}
h(r)\frac{\partial^{2}\psi (r)}{\partial r^{2}}
 + \left[ h'(r) - 2i\omega \right] 
 \frac{\partial\psi (r)}{\partial r}-\tilde{V}(r) \psi(r) = 0 \,\, ,
\label{r_wave_equation}
\end{equation}
where $\tilde{V}(r) = V(r)/h(r) = h'(r)/r+\ell(\ell+1)/r^2$.

Introducing $x=1/r$, Eq.(\ref{r_wave_equation}) can be reexpressed as
Eqs.(15-18) in \cite{Wang-00}. These equations are appropriate to
directly obtain the QN frequencies using the Horowitz-Hubeny method.

%%%%%%%%%%%%%%%%%%%%%%%%%%%%%%%%%%%%%%%%%%%%%%%%%%%%%%%%%%%%%%%%%%%%%%%%%%%%%%%
\subsection{Numerical methods}
%%%%%%%%%%%%%%%%%%%%%%%%%%%%%%%%%%%%%%%%%%%%%%%%%%%%%%%%%%%%%%%%%%%%%%%%%%%%%%%

The quasinormal modes in AdS spacetimes are usually defined as solutions of
the relevant wave equations characterized by purely ingoing waves at
the black hole event horizon and vanishing of the perturbation at
radial infinity. However, we would mention that in the context of the
AdS/CFT correspondence, there is no consensus on the ``correct'' QNM
boundary conditions \cite{Moss}. We will use two different numerical
methods to solve the wave equations.  
 
The first method is the Horowitz-Hubeny method, which has been used
extensively in previous papers \cite{Horowitz-00,Wang-00,Cardoso,Berti-03}. 
It was found in \cite{Wang-00} that with the increase of black hole
charge in RNAdS spacetimes, a large number $N$ of terms in the partial
sum to reduce the relative error in the computation of QN frequencies
is needed. This takes a lot of computer time and cannot easily be performed
for a sum of order $N\geq 300$. We have to use the trial and error
method proposed in \cite{Konoplya-02} to truncate the sum to some
large $N$. However, with big $N$ we adopt, initial tiny error may grow
through recursion relations. We have to improve the precision of all
input data with the help of a built-in function of
\emph{Mathematica}. In addition, high precision of recurrence relation
is also required. After making these improvements, we can get precise
QN frequencies in the lowest lying mode with the increase of black
hole charge and also high overtones.  

However, the Horowitz-Hubeny method breaks down for large values of the
charge for reasons disclosed in \cite{Wang-00}. Time evolution approach do
not suffer the problem and can be counted on for the study of the
highly charged RNAdS black holes. The usual implementation of the
method for two-dimensional d'Alembertians  was introduced in
\cite{Gundlach-94}, and later used in several other contexts
\cite{deSitter_1,Wang-01}. 

It was observed in  \cite{Wang-01} that, for the RNAdS geometries,
this usual method develop minor instabilities with large values of
$r_{+}$ in the form of ``plateaus.'' To overcome this problem  we have
used  another discretization for Eq. (\ref{u_v_wave_equation}),
in addition to the usual one in \cite{Wang-01}, given by   
\begin{multline}
\left[ 1 + \frac{\Delta^{2}}{16}V(S)\right] \psi(N) = 
\psi(E)+\psi(W)-\psi(S) \\
- \frac{\Delta^{2}}{16}
       \left[ V(S)\psi(S)+V(E)\psi(E) \right. \\
+ \left. V(W)\psi(W)\right] \,\,.   
\end{multline}   
The points $N$, $S$, $W$ and $E$ are defined as usual: $N = (u + \Delta, v
+ \Delta)$, $W = (u + \Delta, v)$, $E = (u, v + \Delta)$ and $S =
(u,v)$. The local truncation error is of the order of $O(\Delta^{4})$.
Although this new discretization is more time consuming, we
have observed, on a empirical basis, that it is more stable for fast
decaying fields.  

In presenting the numerical results, we will set the
AdS radius $R=1$ (always) and the black hole radius $r_{+}=100$ (for
the most part). In the
following sections we will first study the properties of the lowest
QNMs for lowly charged black holes, verifying and extending results found in
\cite{Wang-00,Wang-01,Berti-03}. Then we will present our results for lowest
lying modes for highly charged holes. We will also discuss high
overtones of the scalar perturbation for RNAdS black holes and show
the nontrivial angular index influence.  

%%%%%%%%%%%%%%%%%%%%%%%%%%%%%%%%%%%%%%%%%%%%%%%%%%%%%%%%%%%%
\begin{figure}
\resizebox{1\linewidth}{!}{\includegraphics*{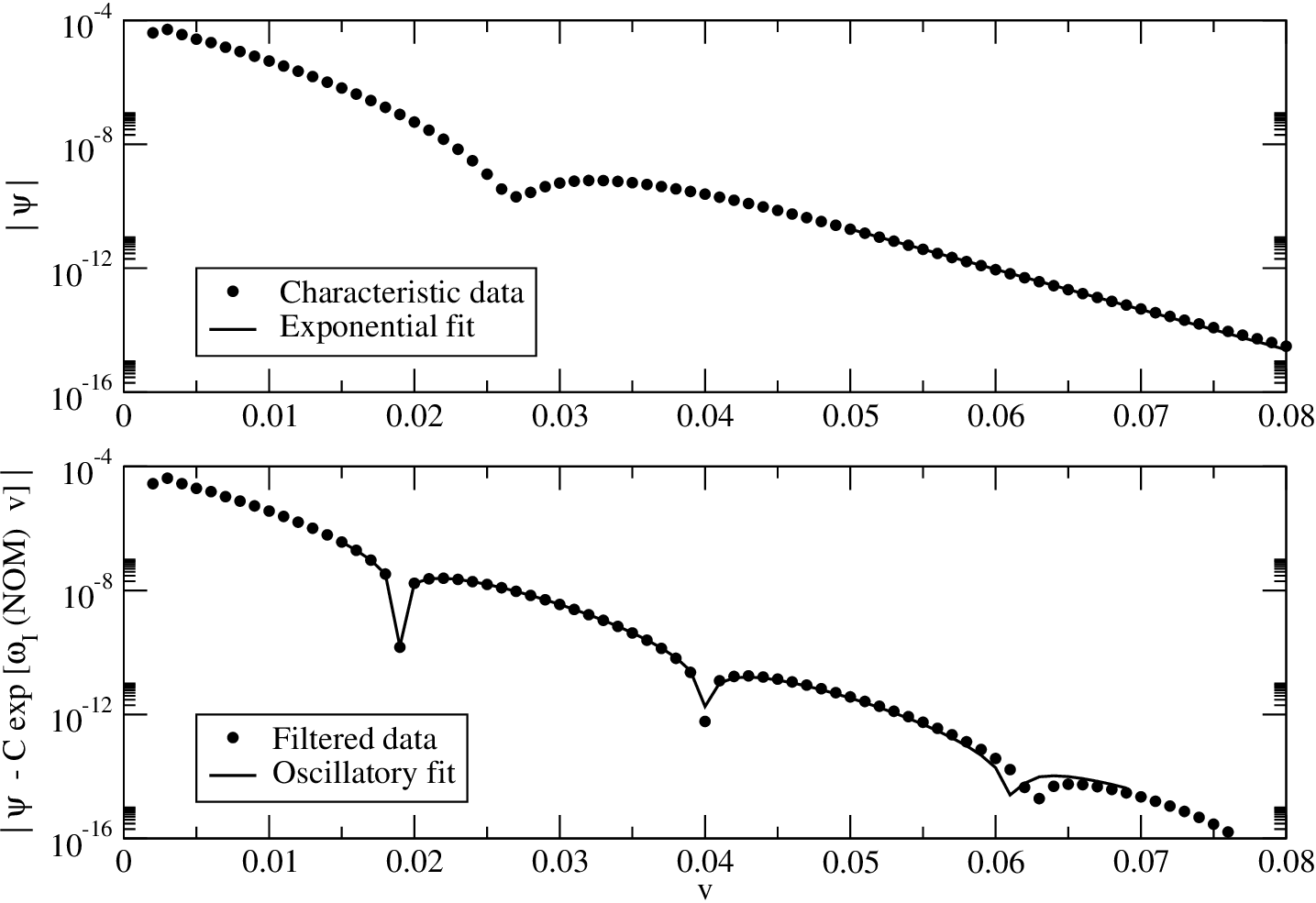}} 
\caption{Illustration of the filtering process. 
  (top) Data obtained from the characteristic
  algorithm. (bottom) The $\chi^{2}$-fitting made in the
  characteristic data subtracted from the Horowitz-Hubeny NOM solution. The
  results obtained from this fitting are compatible with the
  Horowitz-Hubeny OM results. In the graphs
  $r_{+}=100$, $Q/Q_{max} = 0.41$, $R=1$, $\ell=0$ and $n=0$.}     
\label{fig1}
\end{figure}
%%%%%%%%%%%%%%%%%%%%%%%%%%%%%%%%%%%%%%%%%%%%%%%%%%%%%%%%%%%%

%%%%%%%%%%%%%%%%%%%%%%%%%%%%%%%%%%%%%%%%%%%%%%%%%%%%%%%%%%%%%%%%%%%%%%%%%%%%%%%
\section{Numerical results}
%%%%%%%%%%%%%%%%%%%%%%%%%%%%%%%%%%%%%%%%%%%%%%%%%%%%%%%%%%%%%%%%%%%%%%%%%%%%%%%

%%%%%%%%%%%%%%%%%%%%%%%%%%%%%%%%%%%%%%%%%%%%%%%%%%%%%%%%%%%%%%%%%%%%%%%%%%%%%%%
\subsection{Overview of the numerical results}
%%%%%%%%%%%%%%%%%%%%%%%%%%%%%%%%%%%%%%%%%%%%%%%%%%%%%%%%%%%%%%%%%%%%%%%%%%%%%%%

Employing the Horowitz-Hubeny method to the high precision, we
obtained two stable classes of solutions for the perturbation
frequencies. In the first class, oscillation modes (OM) solutions, the
frequencies have both the real and imaginary parts: $\omega
(OM)=\omega_R(OM)+i\omega_I(OM)$, with $\omega_R(OM)\ne0$.  In the second
class of solutions, the nonoscillation modes (NOM) solutions, the
frequencies are purely imaginary: $\omega(NOM)=i\omega_I(NOM)$. In
both cases, the imaginary part of the frequency is always negative,
which indicates that the solutions are stable. 

This scenario is consistent with the results obtained with time
evolution approach. From the results obtained from the characteristic
integration routine, it is possible to estimate with high precision
the oscillatory and exponential decay parameters using a nonlinear fitting
based in a $\chi^2$  analysis. Comparing to the results
obtained by using the time evolution approach, we found that modes
with frequencies choosing by $\min \left\{
\sqrt{\omega_{R}(OM)^{2} + \omega_{I}(OM)^2},
\sqrt{\omega_I(NOM)^2}\right\}$  is consistent with the results
obtained by time evolution approach. We emphasize that the numerical
concordance is excellent. Comparing to the time evolution approach can
help us find the criterion to determine which class of solutions got
by Horowitz-Hubeny method dominates the decay of the scalar field. 
We will use this criterion to order the
modes.  

On the other hand, no matter the criterion used to order the modes,
both classes of solutions are physically relevant.
The characteristic solutions of the wave equation show the asymptotic
behavior of the field propagation. We observe two distinct ``phases''
with the increase of the electric charge. For small $Q$, the scalar
field decays exponentially and oscillates. The frequencies in this
case are compatible with the OM Horowitz-Hubeny solutions. 
For $Q$ greater than some critical value $Q_{c}$, the decay is
purely exponential. The exponential coefficients are  compatible with
the NOM-Horowitz-Hubeny solutions. 

But, even for large  $Q/Q_{max}$, the characteristic approach
does not exclude the OM solutions. It just says that the NOM solution
dominates for large $t$. This does not invalidate the OM solution,
because $\left| \omega_I(NOM) \right| < \left| \omega_{I}(OM) \right|$ and
the oscillatory solutions could be present in the data, but hidden by
the nonoscillatory solutions.    

To expose the OM solution in the characteristic data, we will apply
a filtering procedure: take the wave-function $\Psi(v)$ obtained from the
characteristic integration; subtract from it the exponential part
($\omega_{I}(NOM)$) obtained from Horowitz-Hubeny method:  
\begin{equation}
\Psi(v) \longrightarrow \Psi(v) - C \exp \left[ \omega_{I}(NOM) \, v \right] 
\, \, ;
\end{equation}
and finally check if the remaining part is an oscillatory-exponential
decaying function, compatible with the OM solutions.  

Our results show that both modes are present in the characteristic
results, although for large values of charge, the NOM solutions
dominate the asymptotic limit, as indicated in Table
\ref{table1}. An illustration of the $\chi^2$-fittings made is shown in
Fig. \ref{fig1}.   To simplify notation, we will avoid
the ``$(OM)$'' and ``$(NOM)$'', except where it is not clear from the
context.

\begin{table*}[top]
\caption{Values for the real and imaginary
  frequencies for the fundamental mode ($n=0$), obtained from Horowitz-Hubeny
  method (OM solutions) and filtered characteristic data, for
  $r_{+} = 100$, $R=1$, $\ell=0$.} 
\label{table1}
\begin{ruledtabular}
\begin{tabular}{ccccc}
\multicolumn{1}{c}{}         &
\multicolumn{2}{l}{Horowitz-Hubeny (OM solutions)}      &
\multicolumn{2}{c}{Frequencies after filtering}\\
$Q/Q_{max}$ & $\omega_{R} (OM)$ & $\omega_{I} (OM)$ &
$\omega_{R}$ (error) & $\omega_{I}$  (error) \\ 
\hline 
0.39 & 140.74 & -340.47 & 138.27 (1.75\%)  & -328.98 (3.38\%)    \\
0.40 & 144.54 & -344.62 & 146.96 (1.67\%)  & -355.71( 3.22\%)   \\
0.41 & 147.96 & -347.49 & 148.78 ( 0.55\%) & -347.52  ( 0.009\%) \\
0.42 & 150.77 & -349.57 & 151.10 ( 0.22\%) & -348.96  ( 0.17\%)  \\
0.43 & 152.97 & -351.20 & 153.71 ( 0.48\%) & -350.88  ( 0.09\%)  \\
0.44 & 154.61 & -352.59 & 156.02 ( 0.90\%) &  -352.70  ( 0.03\%) \\
\end{tabular}
\end{ruledtabular}
\end{table*}

%%%%%%%%%%%%%%%%%%%%%%%%%%%%%%%%%%%%%%%%%%%%%%%%%%%%%%%%%%%%
\begin{figure}
\resizebox{1\linewidth}{!}{\includegraphics*{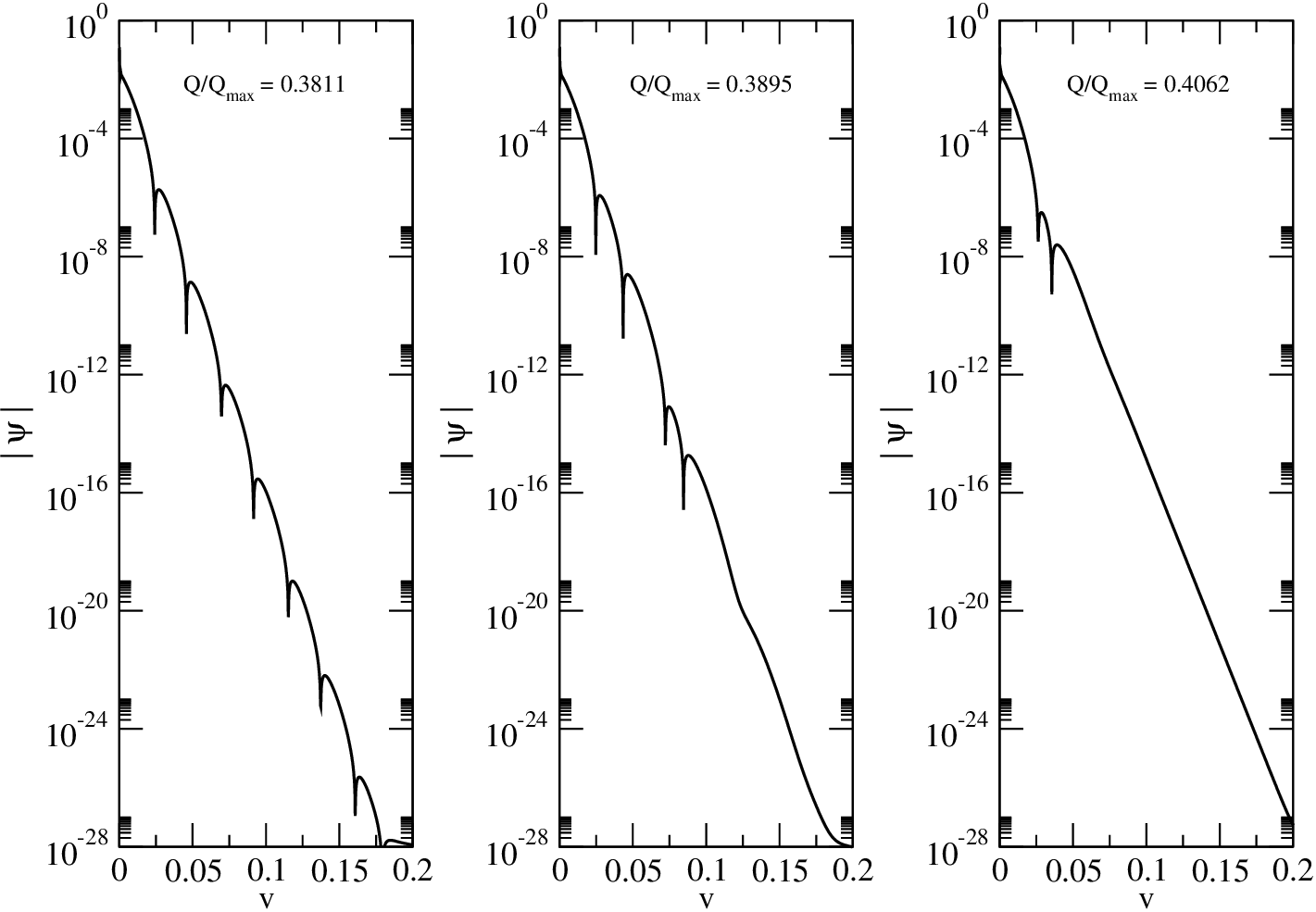}} 
\caption{Semilog graphs of the scalar field in the event horizon,
  showing the transition from oscillatory to nonoscillatory
  asymptotic decay. In the graphs, $r_{+}=100$, $\ell=0$ and $R=1$.}     
\label{fig2}
\end{figure}
%%%%%%%%%%%%%%%%%%%%%%%%%%%%%%%%%%%%%%%%%%%%%%%%%%%%%%%%%%%%

As far as we have observed, the qualitative aspects of the field decay
are independent of the value of the event horizon $r_{+}$. For
completeness, we present in Table \ref{table2} some nonoscillatory
frequencies for  other values of $r_{+}$.

\begin{table*}[top]
\caption{Values for the nonoscillatory frequencies $\omega_{I}(NOM)$
for several values of $r_{+}$ and $\ell=0$,
\mbox{$Q/Q_{max}=0.35$}, $R=1$.}  
\label{table2}
\begin{ruledtabular}
\begin{tabular}{ccccccc}
$r_{+}$ & $\omega_{I}(n=0)$ & $\omega_{I}(n=1)$  & 
$\omega_{I}(n=2)$ & $\omega_{I}(n=3)$ &
$\omega_{I}(n=4)$ & $\omega_{I}(n=5)$  \\ 
\hline 
%
%80  & -137.62 & -316.50 &  -577.77 &   -759.53 &   -913.50 &  -1054.5 \\
%
90  & -229.46 & -663.64 & -1105.3  &  -1547.9  &  -1990.7  &  -2433.5 \\
100 & -437.11 & -1290.6 & -2155.5  &  -3020.5  &  -3885.5  &  -4750.2 \\
110 & -805.32 & -2448.5 & -4093.9  &  -5738.2  &  -7381.9  &  -9025.3 \\
120 & -1486.7 & -4562.5 & -7632.0  & -10698.5  & -13763.5  & -16827.8 
\end{tabular}
\end{ruledtabular}
\end{table*}

%%%%%%%%%%%%%%%%%%%%%%%%%%%%%%%%%%%%%%%%%%%%%%%%%%%%%%%%%%%%%%%%%%%%%%%%%%%%%%%
\subsection{Fundamental modes}
%%%%%%%%%%%%%%%%%%%%%%%%%%%%%%%%%%%%%%%%%%%%%%%%%%%%%%%%%%%%%%%%%%%%%%%%%%%%%%%

\subsubsection{Black holes with small values of charge}

For the black holes with small values of the charge, it was found that
$\omega_I$ increases with $Q$, while $\omega_R$ decreases with $Q$
\cite{Wang-00,Wang-01}. The change of the imaginary part of the frequency
was supported, while the real part of the frequency was found not to have
monotonically decreasing behavior \cite{Berti-03}. We have two complementary
methods and expect to produce a more precise picture of the field's
evolution. 

For black holes with small values of charge, the scalar field decays
exponentially and oscillates. For $Q$ greater than a critical value,
the decay becomes purely exponential. This point can be directly seen
in the wave function plotting from Fig.\ref{fig2}.
We read from Fig.\ref{fig2} that the oscillation starts to
disappear at $Q_c=0.3895 Q_{max}$ (the first several rings in the plot
$Q_c=0.3895 Q_{max}$ are due to initial pulse). 

%%%%%%%%%%%%%%%%%%%%%%%%%%%%%%%%%%%%%%%%%%%%%%%%%%%%%%%%%%%%
\begin{figure}
\resizebox{1\linewidth}{!}{\includegraphics*{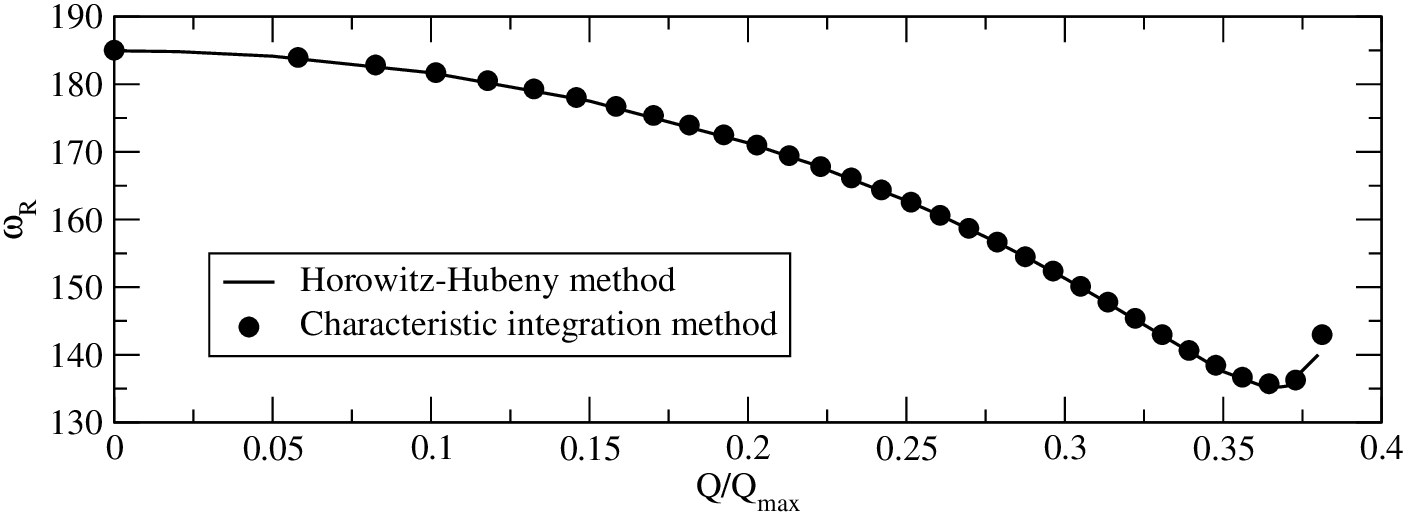}} 
\caption{Graph of $\omega_{R}(OM)$ with $Q/Q_{max}$, using
  Horowitz-Hubeny method and  characteristic integration results. For
  this graph, $r_{+}=100$, $R=1$, $\ell=0$ and $n=0$.}     
\label{fig3}
\end{figure}
%%%%%%%%%%%%%%%%%%%%%%%%%%%%%%%%%%%%%%%%%%%%%%%%%%%%%%%%%%%%

For $Q<Q_c$ the scalar perturbation experiences oscillation. The behavior of the
real part of QN frequencies with the increase of the charge is shown
in Fig.\ref{fig3}, where results obtained from Horowitz-Hubeny
and characteristic integration methods are presented. 

%%%%%%%%%%%%%%%%%%%%%%%%%%%%%%%%%%%%%%%%%%%%%%%%%%%%%%%%%%%%
\begin{figure}
\resizebox{1\linewidth}{!}{\includegraphics*{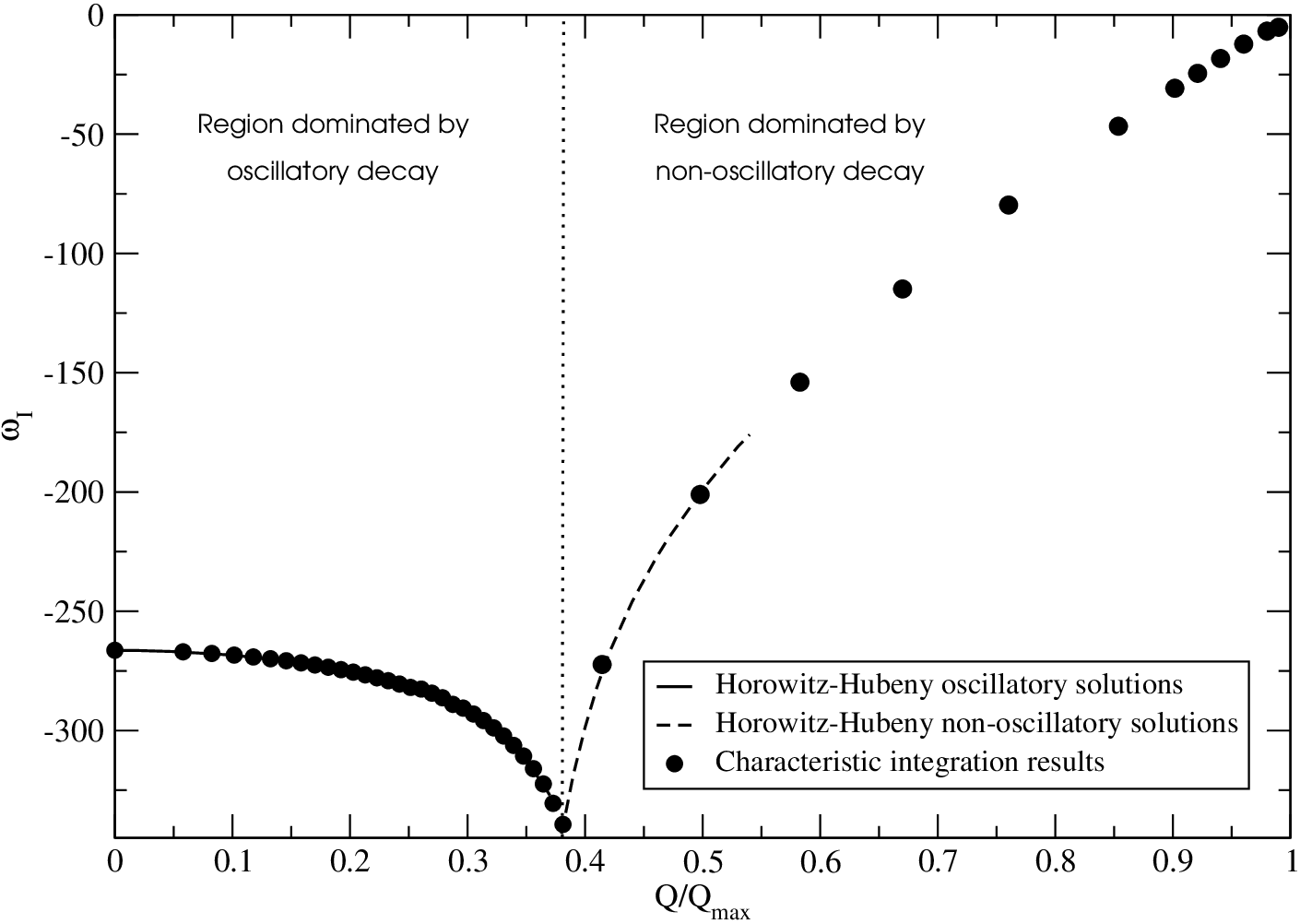}} 
\caption{Graph of  $\omega_{I}$ (OM and NOM) with $Q/Q_{max}$, showing that
  $\omega_{I}(NOM)$ tends to zero as $Q$ tends to $Q_{max}$. In the
  graph, $r_{+}=100$, $R=1$, $\ell=0$ and $n=0$.}     
\label{fig4}
\end{figure}
%%%%%%%%%%%%%%%%%%%%%%%%%%%%%%%%%%%%%%%%%%%%%%%%%%%%%%%%%%%%

It is apparent that the  different numerical methods have very good
agreement. We found indeed the minimum of $\omega_R$ at a certain
value of charge, $Q_e=0.366Q_{max}$ from the Horowitz-Hubeny method
and $Q_e=0.3645Q_{max}$ from the characteristic method. This result
confirms that observed in \cite{Berti-03}. After the minima of
$\omega_R$, $\omega_R$ increases with $Q$ a bit, but not as much as
described in \cite{Berti-03}. Starting from $Q_c$, the NOM starts to dominate. The real part of the frequency of the perturbation 
vanishes and will not reach a local maxima at $Q=0.474Q_{max}$ as obtained
in \cite{Berti-03}.   

Now let us turn to discuss the imaginary part of the QN
frequencies. Our two numerical methods give consistent results. From
Fig.\ref{fig4}, we found before $Q_e$ where $\omega_R$ has minima, $\vert
\omega_I\vert$ monotonically increases with charge. This result agrees
to that obtained in \cite{Wang-00,Berti-03}. However, there is no change of
sign of the second derivative of $\omega_I(Q/Q_{max})$ as observed in
\cite{Berti-03}.    $\vert \omega_I\vert$ reaches its maxima at the same
value of the charge $Q_e$ as $\omega_R$ arrives at its minima. For
$Q<Q_e$, our previous statement is still valid, ``If we perturb a
RNAdS black hole with high charge, the surrounding geometry will not
`ring' as much and long as that of the black hole with small
charge.'' \cite{Wang-00,Wang-01}. For $Q>Q_e$, Fig.4 tells us that $\vert
\omega_I\vert$ decreases with the increase of charge until
$Q=Q_{max}$. The turning point of $\omega_I$ appears at a value of $Q$ that is not very big, and it is different from the rough argument in \cite{Wang-01}. 

%%%%%%%%%%%%%%%%%%%%%%%%%%%%%%%%%%%%%%%%%%%%%%%%%%%%%%%%%%%
\begin{figure}
\resizebox{1\linewidth}{!}{\includegraphics*{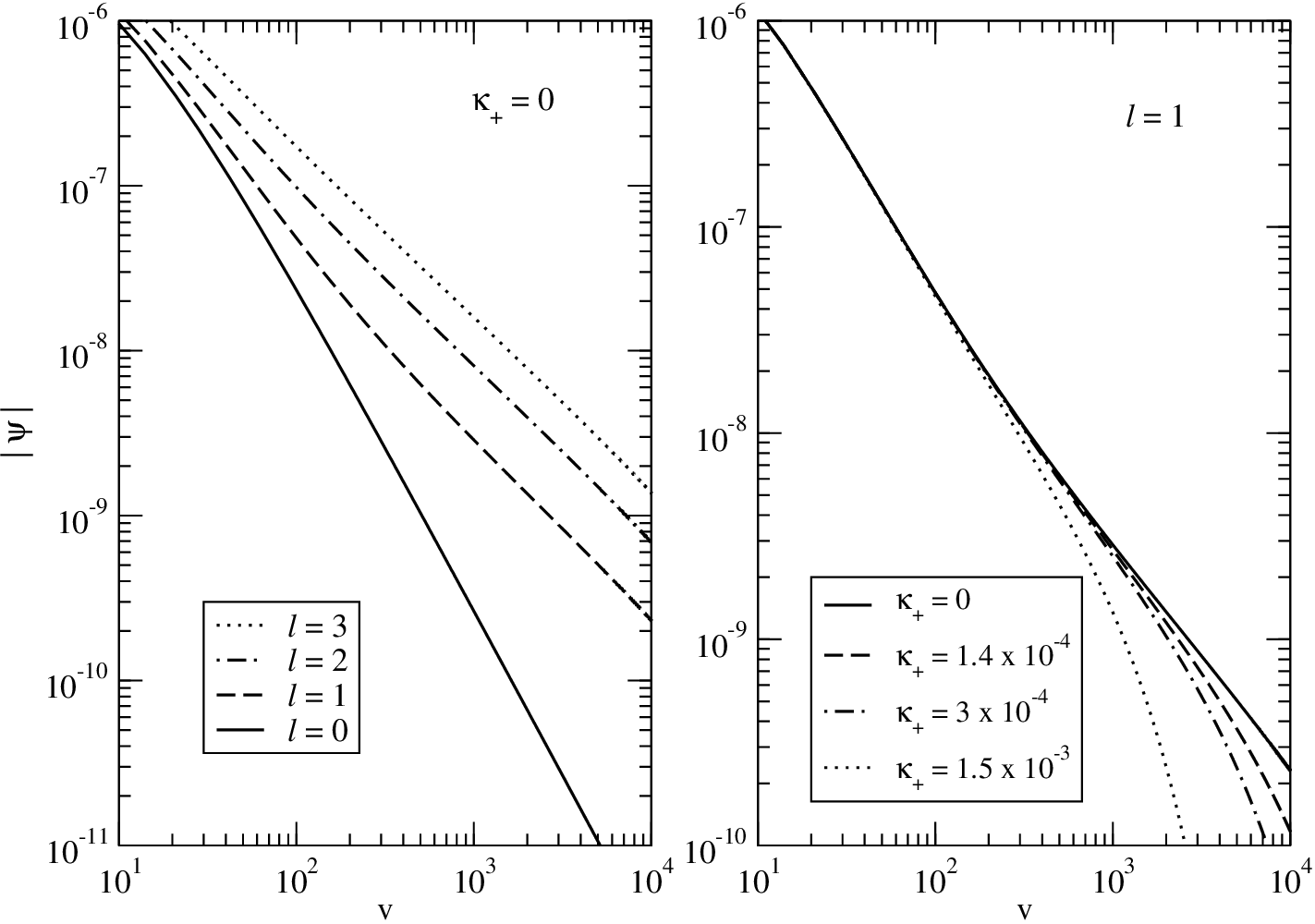}} 
\caption{Log-log graphs of the scalar field in the event
  horizon. (left) Extreme case for different values of $\ell$, showing
  power-law tails. If $\ell = 0$, $\Psi \propto v^{-2}$, but if $\ell >
  0$, $\Psi \propto v^{-1}$. (right) Approach to the extreme limit,
  showing transition from exponential decay to power-law decay
  ($\ell=1$). For all curves, $r_{+}=100$ and $R=1$.}   
\label{fig5}    
\end{figure}
%%%%%%%%%%%%%%%%%%%%%%%%%%%%%%%%%%%%%%%%%%%%%%%%%%%%%%%%%%%%

\subsubsection{Extreme and near extreme limits}

From Fig.\ref{fig3} and Fig.\ref{fig4} we got the picture that for
$Q>Q_c$ the asymptotic scalar perturbation is dominated by the
nonoscillatory modes, while $|\omega_{I}|$ keeps decreasing
monotonically and smoothly up to $Q=Q_{max}$. We learned
from Fig.\ref{fig4} that as the black hole becomes extreme,
$\omega_{I}$ tends to zero, which seems to confirm the conjecture
raised in Section 3.3.2 of \cite{Berti-03}.    

However, from the tail behavior (Fig.\ref{fig5}) we
observed that in the extreme limit, the decay of scalar perturbation
still exists, though changing from the exponential to power-law
decay. This result strongly suggests that due to scalar perturbations
the extreme RNAdS black hole is stable, not marginally unstable as
worried in \cite{Berti-03} by considering electromagnetic and axial
gravitational perturbations.   

The reason of the smooth change of the perturbation behavior can
be understood as follows. In the extreme limit, the horizon $r_{+}$ is
a double root of $h(r)$, and therefore the expression for the tortoise
coordinate is given by 
\begin{multline}
r_{ext}^{*}(r)  = 
2 A_{ext}\ln\left(r-r_{+}\right) - \frac{C}{r-r_{+}} 
- A_{ext} \ln(r^{2}+p_{ext}r \\ 
+q_{ext}) 
+ \frac{2(B_{ext} + Ap_{ext})}{\sqrt{4q_{ext}-p_{ext}^{2}}}
\left[\arctan
  \left(\frac{2r+p_{ext}}{\sqrt{4q_{ext}-p_{ext}^{2}}}\right)
 -\frac{\pi}{2}\right] \\
\label{ext_tortoise}
\end{multline}
with the constants $A_{ext}$, $B_{ext}$, $C$, $p_{ext}$ and $q_{ext}$
defined as 
\begin{equation}
A_{ext}  = \frac{R^{2} r_{+}\left(4r_{+}^{2}
  + R^{2}\right)}{\left(6r_{+}^{2} + R^{2}\right)^{2}} \,\, ,
\end{equation}
\begin{equation}
B_{ext}  =  \frac{R^{2} (6r_{+}^{4} + 5r_{+}^{2}R^{2} +
  R^{4})}{\left(6r_{+}^{2} + R^{2}\right)^{2}}  \,\, ,
\end{equation}
\begin{equation}
C = \frac{R^{2} r_{+}^{2}}{6r_{+}^{2}+R^{2}}  \,\, ,
\end{equation}
\begin{equation}
p_{ext} = 2r_{+}  \,\, , \,\,q_{ext} = R^{2}+3r_{+}^{2} \,\, .
\end{equation}

Globally, the structure of the extreme space-time changes drastically
from that of the nonextreme \cite{Wang_Abdalla_Su}. However, we will
argue that, restricted to the exterior of the black hole, the
transition from near extreme to extreme is, in a sense, smooth. To
parametrize the approach to the extreme limit, we introduce the
dimensionless variable     
\begin{equation}
\delta=\frac{r_{+}-r_{-}}{r_{+}} \,\, .
\end{equation}
In terms of $\delta$, the near extreme limit is given by $0 <
\delta\ll 1$, and the extreme case is $\delta = 0$.

Let us analyze the two first terms in the expression for the tortoise
function Eq.(\ref{ne_tortoise}). We will expand these expressions in
$\delta$. For this, we note that
\begin{widetext}

\begin{equation}
\frac{1}{2\kappa_{+}}\ln\left(r-r_{+}\right) - \frac{1}{2\kappa_{-}}
\ln\left(r-r_{-}\right) =
-\left[ \frac{1}{\delta} \frac{R^{2}r_{+}}{6r_{+}^{2}+R^{2}} -
 \frac{2R^{2}r_{+} \left(2r_{+}^{2} + R^{2}\right)}{\left(6r_{+}^{2} +
 R^{2}\right)^{2}}\right] \ln\left(1 + \frac{r_{+}}{r -
 r_{+}}\delta\right) + O\left(\delta\right) \,\, .
\end{equation}

At this point, it is necessary to consider another limit. If $r$
is not too close to the horizon, or more precisely, if $r_{+} \delta
\ll r-r_{+}$ 
then
\begin{equation}
\frac{1}{2\kappa_{+}}\ln\left(r-r_{+}\right) - \frac{1}{2\kappa_{-}}
\ln\left(r-r_{-}\right) =
\frac{2R^{2}r_{+} \left(4r_{+}^{2} + R^{2}\right)}{\left(6r_{+}^{2} +
  R^{2} \right)^{2}} \ln \left(r-r_{+}\right) -
  \left(\frac{R^{2}r_{+}^{2}}{6r_{+}^{2} + R^{2}}\right)
  \frac{1}{r-r_{+}} + O\left(\delta\right)
\end{equation}
On the other hand, for $r$ sufficiently close to $r_{+}$, or more
specifically if $r_{+} \delta \gg r-r_{+}$:
\begin{equation}
\frac{1}{2\kappa_{+}} \ln\left(r-r_{+}\right) - \frac{1}{2\kappa_{-}}
\ln \left(r-r_{-}\right) \approx
\frac{1}{\delta} \frac{R^{2}r_{+}}{6r_{+}^{2}+R^{2}}
\ln\left(r-r_{+}\right) 
\end{equation}

\end{widetext}

The near extreme approximations for the other terms in the tortoise
functions are simpler, one can substitute $A \rightarrow A_{ext}$,
$B \rightarrow B_{ext}$, $p \rightarrow p_{ext}$ and  $q \rightarrow
q_{ext}$, with an error of the order of $\delta$. 
Using the previous
results, we find that in the near extreme limit we have 
\begin{equation}
r^*(r) = \left\{ \begin{array}{ccc}
  r^*_{ext}(r) + O\left(\delta\right)& \textrm{if} &
  \frac{r_{+}}{r-r_{+}} \delta \ll 1 \\ 
\frac{1}{\delta} \frac{R^{2}r_{+}}{6r_{+}^{2}+R^{2}} \ln\left(r -
r_{+} \right) + \cdots & \textrm{if} & \frac{r_{+}}{r - r_{+}} \delta
\gg1 
                \end{array}\right. \,\, .
\end{equation}
We see that for a certain value of $r$, the tortoise function
becomes dominated by the power-law term, characteristic of the extreme
function. When very near the horizon, the function $r^*(r)$ is dominated
by a logarithmic term, which diverges when $r\rightarrow r_+$.

%%%%%%%%%%%%%%%%%%%%%%%%%%%%%%%%%%%%%%%%%%%%%%%%%%%%%%%%%%%%
\begin{figure}
\resizebox{1\linewidth}{!}{\includegraphics*{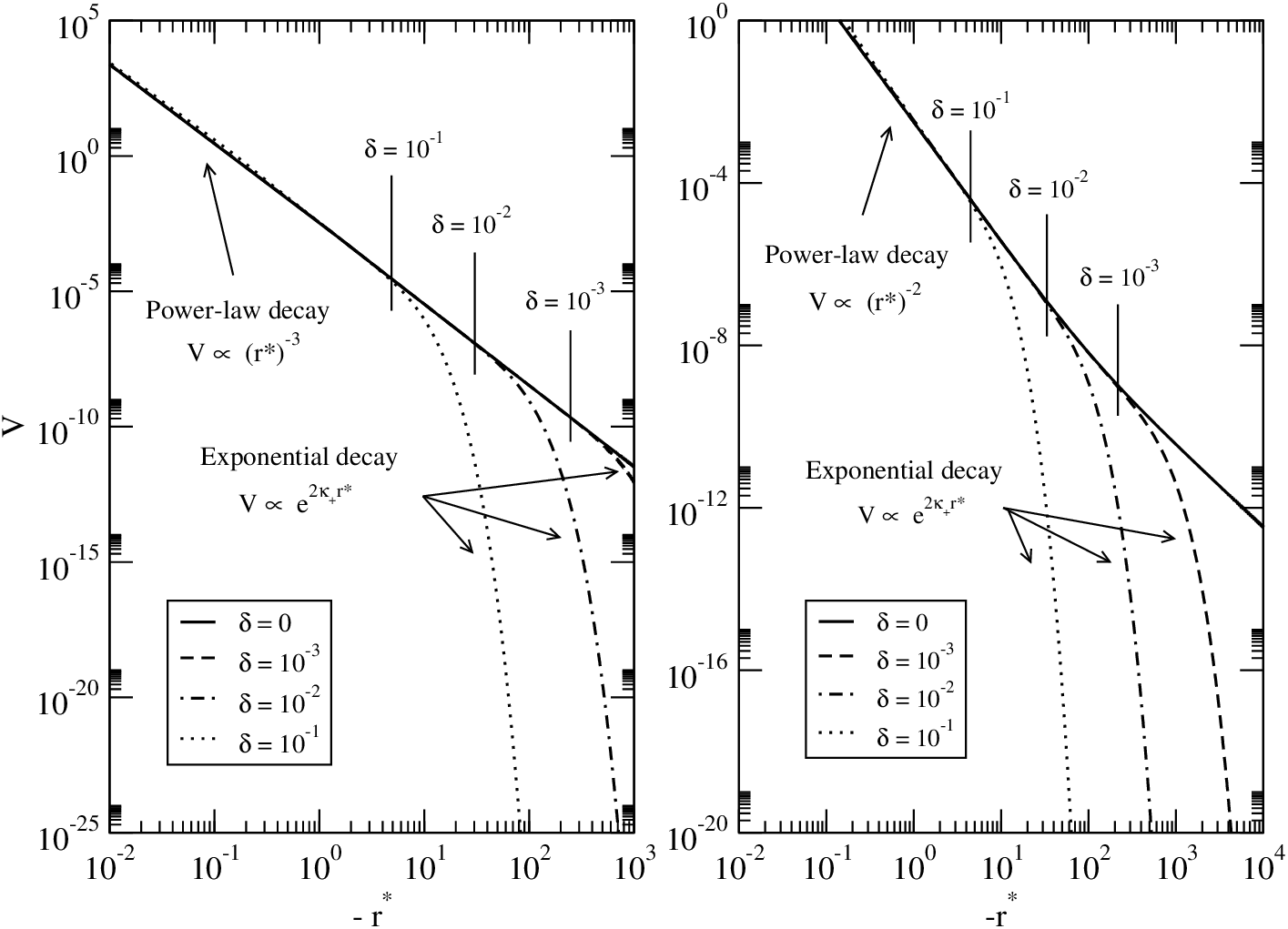}} 
\caption{Log-log graphs of the function $V(r^{*})$,
  from the near extreme limit approaching the extreme limit. At a
  certain point, if $\delta > 0$, there is a transition from power-law
  decay to exponential decay, indicated for each graph. For the curves,
  $r_{+}=100$, $R=1$ and $\ell=0,1$ (left and right figures,
  respectively).}  
\label{fig6}    
\end{figure}
%%%%%%%%%%%%%%%%%%%%%%%%%%%%%%%%%%%%%%%%%%%%%%%%%%%%%%%%%%%%

For the effective scalar potential, using the previous results, we
find that, if $r_{+} \delta \ll r-r_{+}$ , including the extreme limit
where $\delta = 0$:
\begin{equation}
V(r^{*}) \propto \left\{ \begin{array}{ccc}
   (r^{*})^{-2} & \textrm{if} & \ell > 0 \\
   (r^{*})^{-3} & \textrm{if} & \ell=0
       \end{array}\right. \,\, .
\label{pot_near_ext}
\end{equation}
The difference between $\ell=0$ and $\ell>0$ in expression
(\ref{pot_near_ext}) comes from the presence of a non-null term $\ell
(\ell + 1)/r^{2}$ in the potential. This is reflected in
the asymptotic behavior of the wave functions, as seen in
Fig.\ref{fig5}(left). 
 
If $\delta > 0$, the behavior of the effective potential can be
different if the observation point is sufficiently close to the event
horizon. More precisely, if $r - r_{+} \ll \delta r_{+}$, the
exponential decay dominates, and
\begin{equation}
V(r^{*}) \propto e^{2\kappa_{+} r^{*}} \,\, .
\end{equation}

The different potential asymptotic forms, when $Q$ is close to $Q_{max}$,
depend on whether we are initially near to the black hole horizon or
not. This difference will disappear in the black hole extreme limit
($\delta \rightarrow 0$) as illustrated in Fig.6. As
$\delta$ approaches zero, the potential is more and
more dominated by the power-law phase, and the
exponential tail is less relevant. In the extreme limit, there is only a power-law
tail. This smooth change in the RNAdS potential explains the smooth
change in the wave functions from the near extreme to extreme limit,
illustrated in Fig.\ref{fig5}(right).

%%%%%%%%%%%%%%%%%%%%%%%%%%%%%%%%%%%%%%%%%%%%%%%%%%%%%%%%%%%%%%%%%%%%%%%%%%%%%%%
\subsection{Higher modes}
%%%%%%%%%%%%%%%%%%%%%%%%%%%%%%%%%%%%%%%%%%%%%%%%%%%%%%%%%%%%%%%%%%%%%%%%%%%%%%%

The first numerical study of the high overtone QN frequencies for
scalar perturbation was done in SAdS black hole
\cite{Cardoso_Lemos}. We have extended their
investigation to RNAdS spacetimes and found that ``switching on the
charge'' brings interesting physics. For the higher modes, the time
evolution method is not practical. However the Horowitz-Hubeny method
is still efficient, at least for values of the charge not too big.   

For the same value of the charge, both real and imaginary part of QN
frequencies increases with the overtone number $n$. This result
confirms that observed in \cite{Cardoso_Lemos}.  

\begin{figure}
\resizebox{1\linewidth}{!}{\includegraphics*{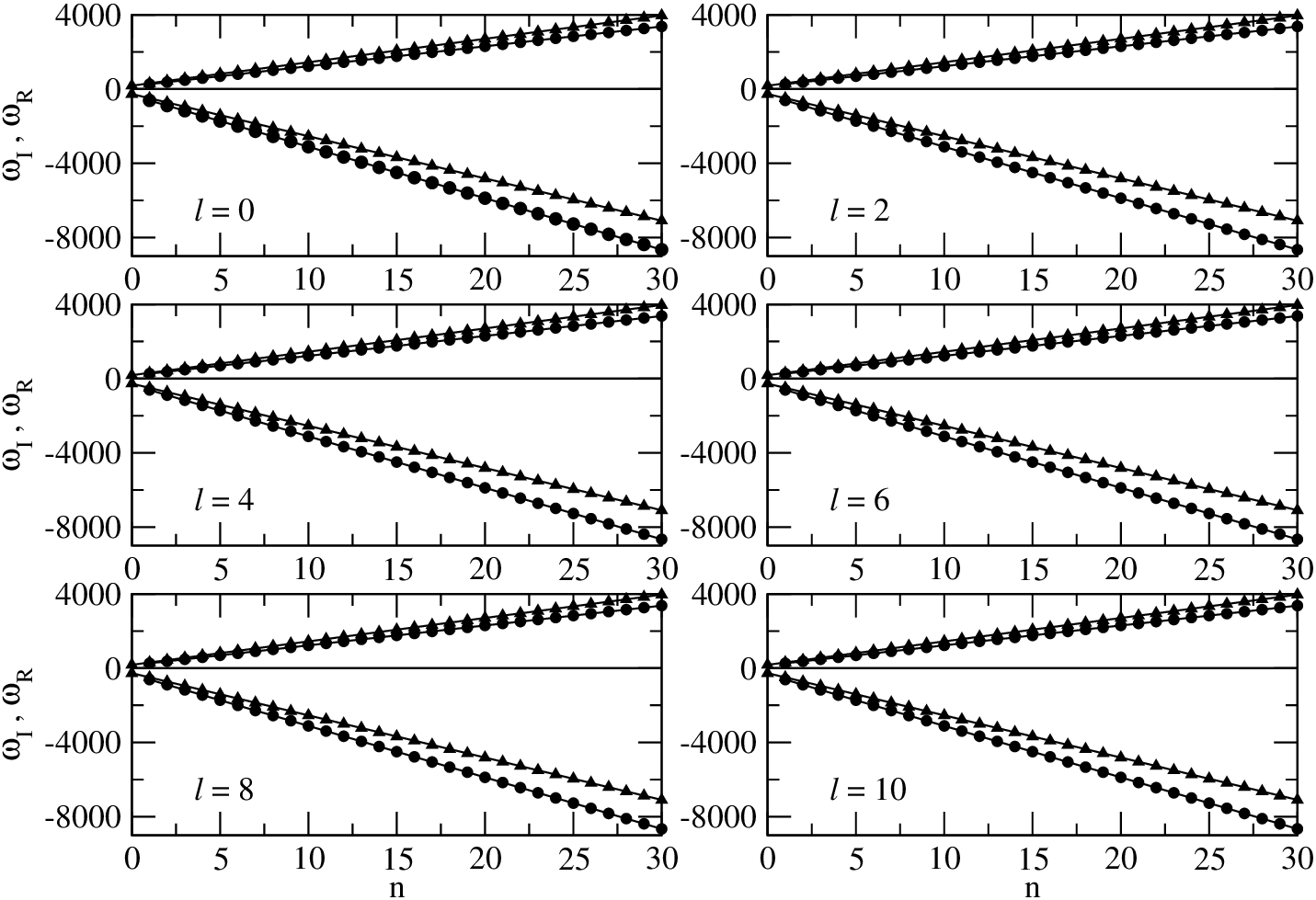}} 
\caption{Graphs of $\omega_{R}(OM)$ and
  $\omega_{I}(OM)$ versus $n$, for $Q/Q_{max}=0.1$ (represented by triangles)
  and $Q/Q_{max}=0.4$ (represented by bullets). The continuous lines
  indicate the linear fits. With the increase of Q, $\omega_{R}(OM)$
  increases slower with $n$, while  $|\omega_{I}(OM)|$ increases
  faster with $n$. For the curves, $r_{+}=100$, $R=1$ and
  $\ell=0,2,4,6,8,10$.}   
\label{fig7}    
\end{figure}

There are large variations of both $\omega_{R}$ and
$\omega_{I}$ with the charge and the overtone
number. Figure \ref{fig7} tells us that with the increase of $Q$,
$\omega_{R}$ increases slower while $| \omega_{I} |$ increases
faster for bigger overtone number $n$.  
%
%
%%%%%%%%%%%%%%%%%%%%%%%%%%%%%%%%%%%%%%%%%%%%%%%%%%%%%%%%%%%%
%%%%%%%%%%%%%%%%%%%%%%%%%%%%%%%%%%%%%%%%%%%%%%%%%%%%%%%%%%%%

\begin{figure}
\resizebox{1\linewidth}{!}{\includegraphics*{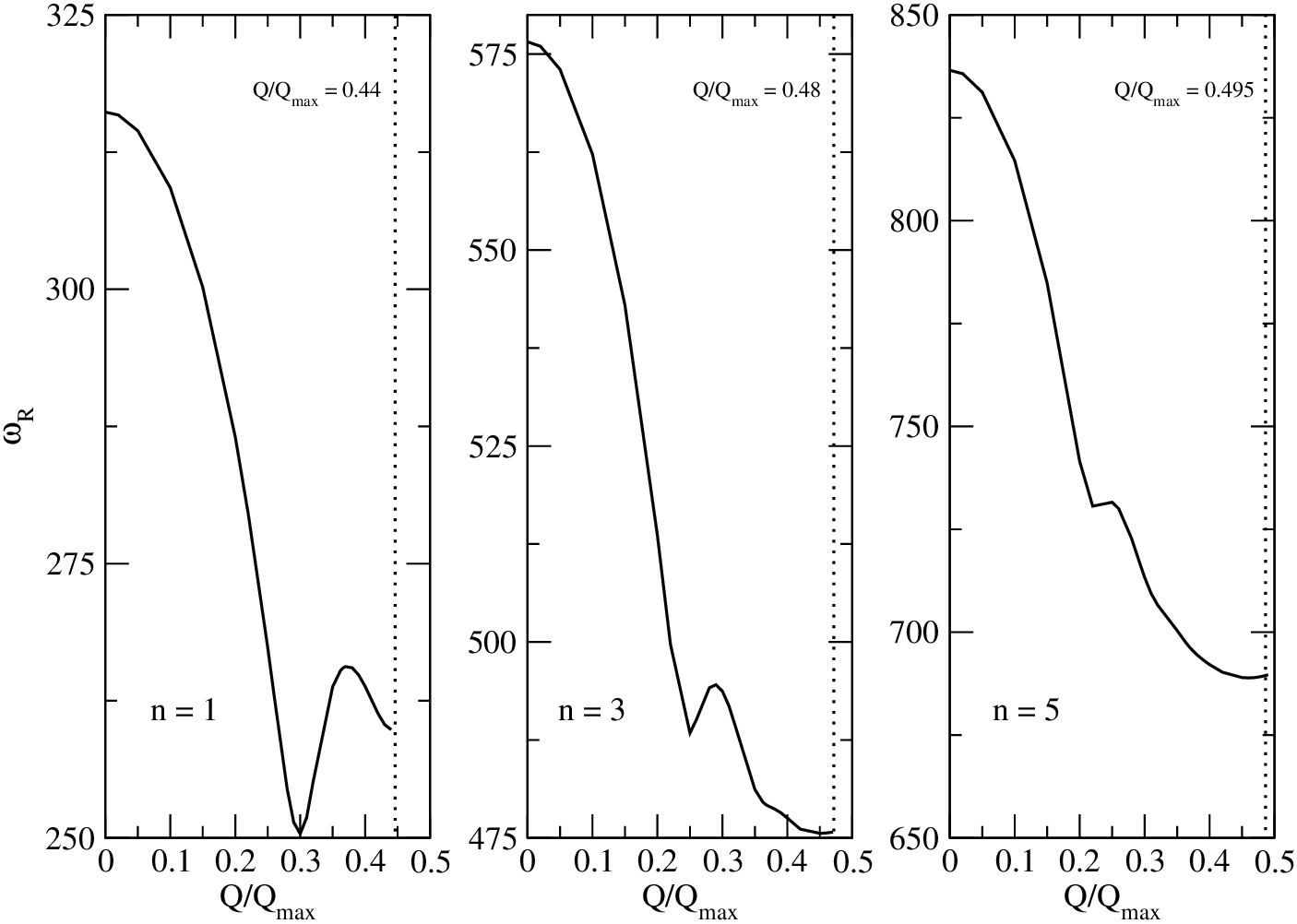}} 
\caption{Graphs of $\omega_{R} (OM)$ with $Q/Q_{max}$,
  showing the ``wiggle behavior''. Values of $Q/Q_{max}$ in graphs
  indicating when nonoscillating solution dominates.
  For the curves, $r_{+}=100$ and $\ell=0$.}  
\label{fig8}    
\end{figure}

\begin{figure}
\resizebox{1\linewidth}{!}{\includegraphics*{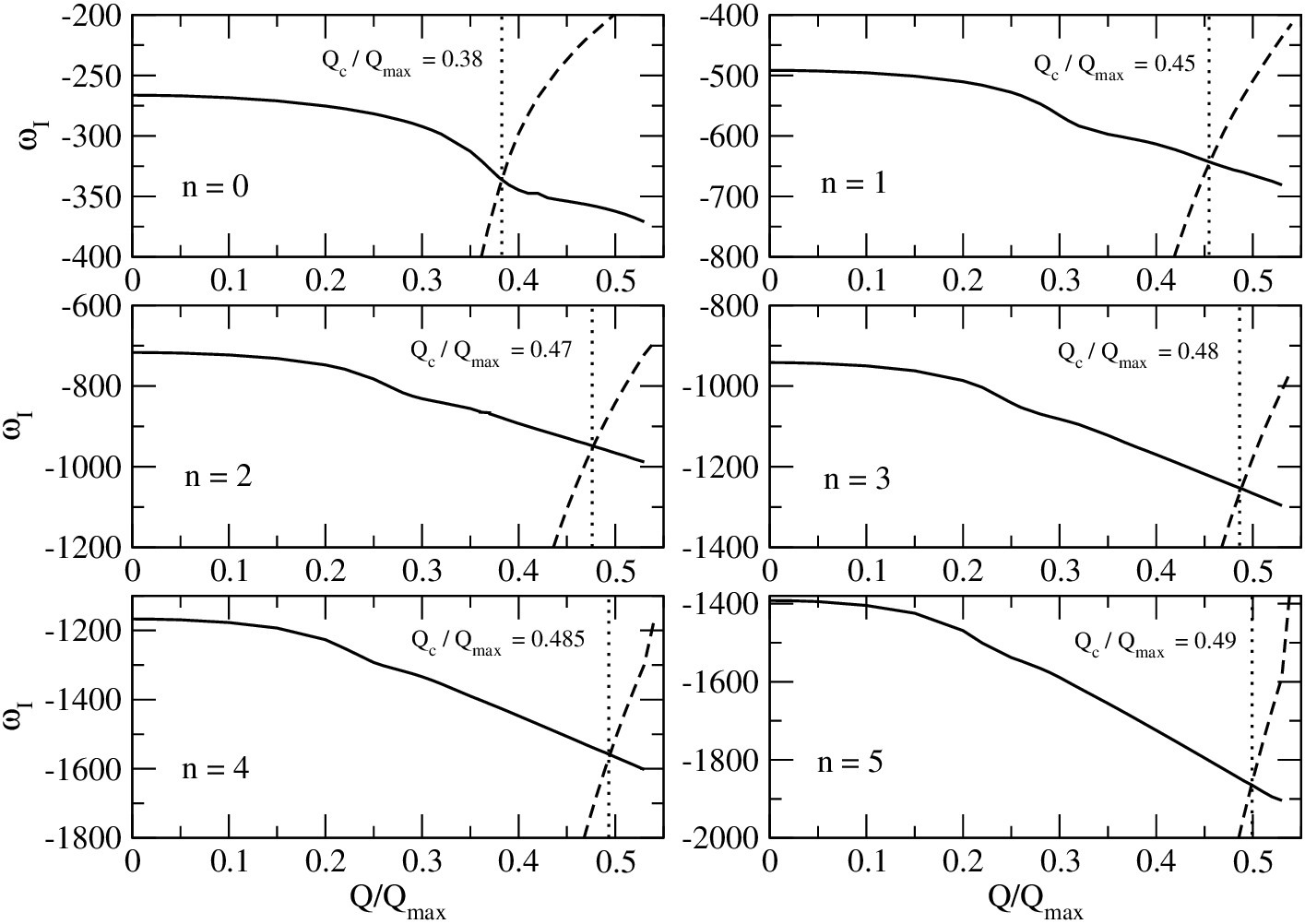}} 
\caption{Graphs of $\omega_{I}$ (OM and NOM solutions) with
  $Q/Q_{max}$ for several values of $n$. The continuous lines indicate
  the oscillatory solution, dotted lines indicate nonoscillatory solutions.
  For the curves, $r_{+}=100$, $R=1$ and $\ell=0$.}  
\label{fig9}    
\end{figure}

\begin{figure}
\resizebox{1\linewidth}{!}{\includegraphics*{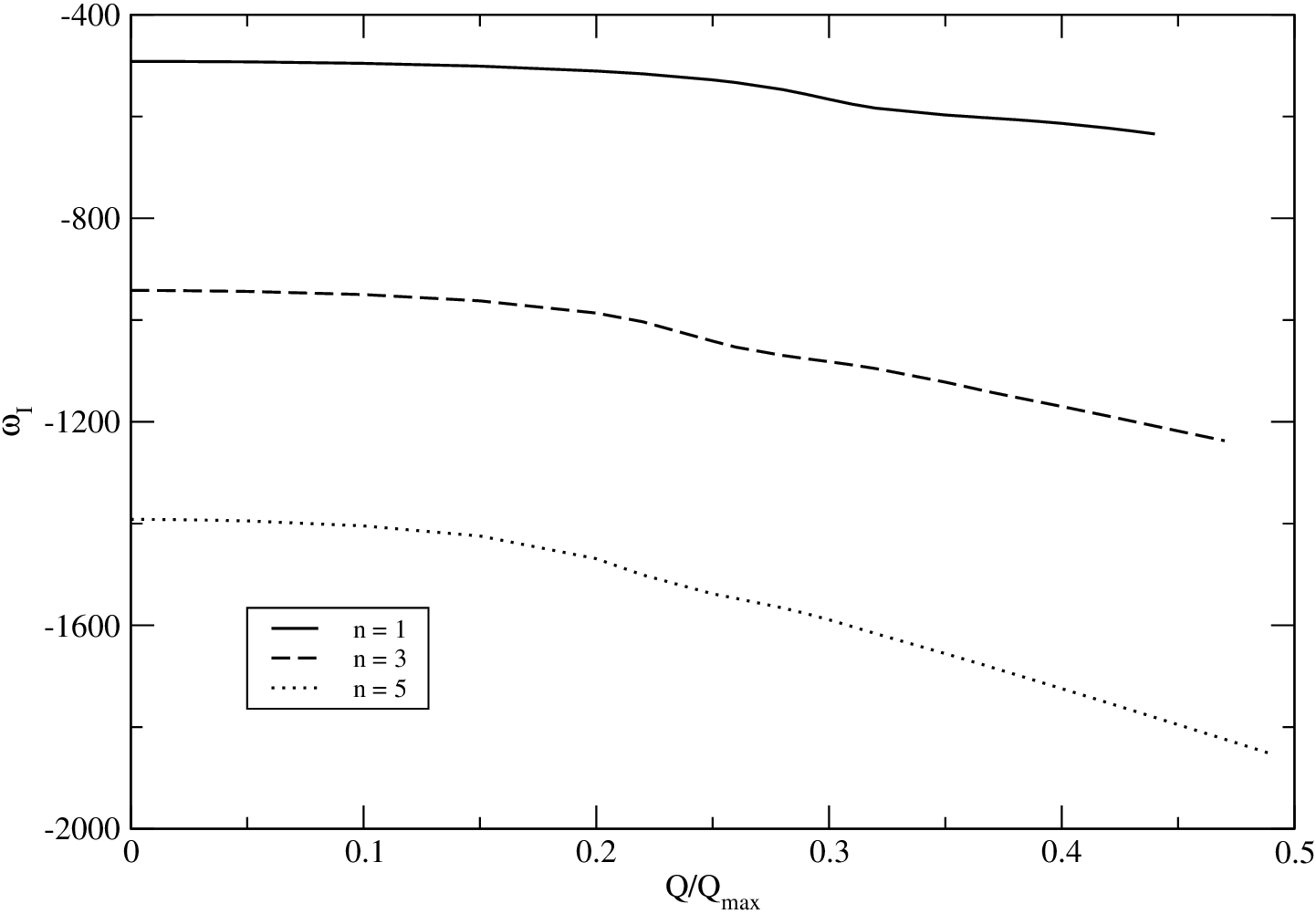}} 
\caption{Graphs of $\omega_{I} (OM)$ with $Q/Q_{max}$, showing the
  increase of  $|\rm{Im}(\omega)|$ with  $n$.  For the curves,
 $r_{+}=100$, $R=1$, $\ell=0$ and $n=0$.} 
\label{fig10}    
\end{figure}

For $\omega_R$, combining Fig.\ref{fig3} and Fig.\ref{fig8}, we see that
the local minima of the ``wiggle'' appear at smaller values of the
charge with the increase of $n$.  The ``wiggle'' of $\omega_R$ becomes
shallower with the increase of $n$ and disappears for $n\geq 5$ as
exhibited in Fig.\ref{fig8}. Further with the increase of $n$, we
observed that $\omega_R$ starts to be zero at bigger values of
charge. For the imaginary part of the QN frequencies, the relation
with $n$ and $Q$ are shown in Figs.\ref{fig9} and \ref{fig10}. We found
that with the increase of $n$, the maxima of $\vert \omega_I\vert$
appear at bigger values of black hole charge. Moreover we saw from
Fig.\ref{fig10} that before the maxima, $\vert \omega_I\vert$
increases quicker with charge for bigger $n$. These richer physics are
brought by introducing the additional parameter $Q$ in RNAdS
spacetimes.

%%%%%%%%%%%%%%%%%%%%%%%%%%%%%%%%%%%%%%%%%%%%%%%%%%%%%%%%%%%%
%%%%%%%%%%%%%%%%%%%%%%%%%%%%%%%%%%%%%%%%%%%%%%%%%%%%%%%%%%%%

%%%%%%%%%%%%%%%%%%%%%%%%%%%%%%%%%%%%%%%%%%%%%%%%%%%%%%%%%%%%
%%%%%%%%%%%%%%%%%%%%%%%%%%%%%%%%%%%%%%%%%%%%%%%%%%%%%%%%%%%%

%%%%%%%%%%%%%%%%%%%%%%%%%%%%%%%%%%%%%%%%%%%%%%%%%%%%%%%%%%%%
%%%%%%%%%%%%%%%%%%%%%%%%%%%%%%%%%%%%%%%%%%%%%%%%%%%%%%%%%%%%

In \cite{Cardoso_Konoplya_Lemos} it was interestingly found
that in the large black hole regime the frequencies become evenly
spaced for high overtone number $n$. For lowly charged RNAdS black
hole, our result confirmed their argument for fixed value of the
charge. The spacing between frequencies are  
\begin{equation}
\omega_{n+1}-\omega_{n} = 129.9-225i, \hspace{1cm} (n\rightarrow
\infty \,\, \textrm{when} \,\, Q=0), 
\end{equation}
which is the same as (12) in \cite{Cardoso_Konoplya_Lemos} by
choosing $r_+=100$ there, and 

\begin{gather}
\omega_{n+1}-\omega_{n}=120.6-233i \nonumber \\
(n\rightarrow
\infty  \,\, \textrm{when} \,\, Q=0.15Q_{max}), 
\end{gather}

\begin{gather}
\omega_{n+1}-\omega_{n}=111.1-256i, \nonumber \\
(n\rightarrow
\infty   \,\, \textrm{when} \,\, Q=0.3Q_{max}). 
\end{gather}
We found that choosing bigger values of the charge, the real part in
the spacing expression becomes smaller, while the imaginary part
becomes bigger.

%%%%%%%%%%%%%%%%%%%%%%%%%%%%%%%%%%%%%%%%%%%%%%%%%%%%%%%%%%%%%%%%%%%%%%%%%%%%%%%
\subsection{Dependence on the angular index $\ell$}
%%%%%%%%%%%%%%%%%%%%%%%%%%%%%%%%%%%%%%%%%%%%%%%%%%%%%%%%%%%%%%%%%%%%%%%%%%%%%%%

We have so far discussed only the QNMs with $\ell=0$. Increasing
$\ell$, for the lowest lying mode, we obtained the effect of
decreasing of $\omega_I$ and increasing of $\omega_R$ in RNAdS
spacetimes \cite{Wang-00,Wang-01}, though the dependence is very weak
\cite{Berti-03}. In studying the higher modes, it was argued that the
asymptotic behavior of the spacing of high overtones is independent of
the value of $\ell$ \cite{Cardoso_Konoplya_Lemos}.  

\begin{figure}
\resizebox{1\linewidth}{!}{\includegraphics*{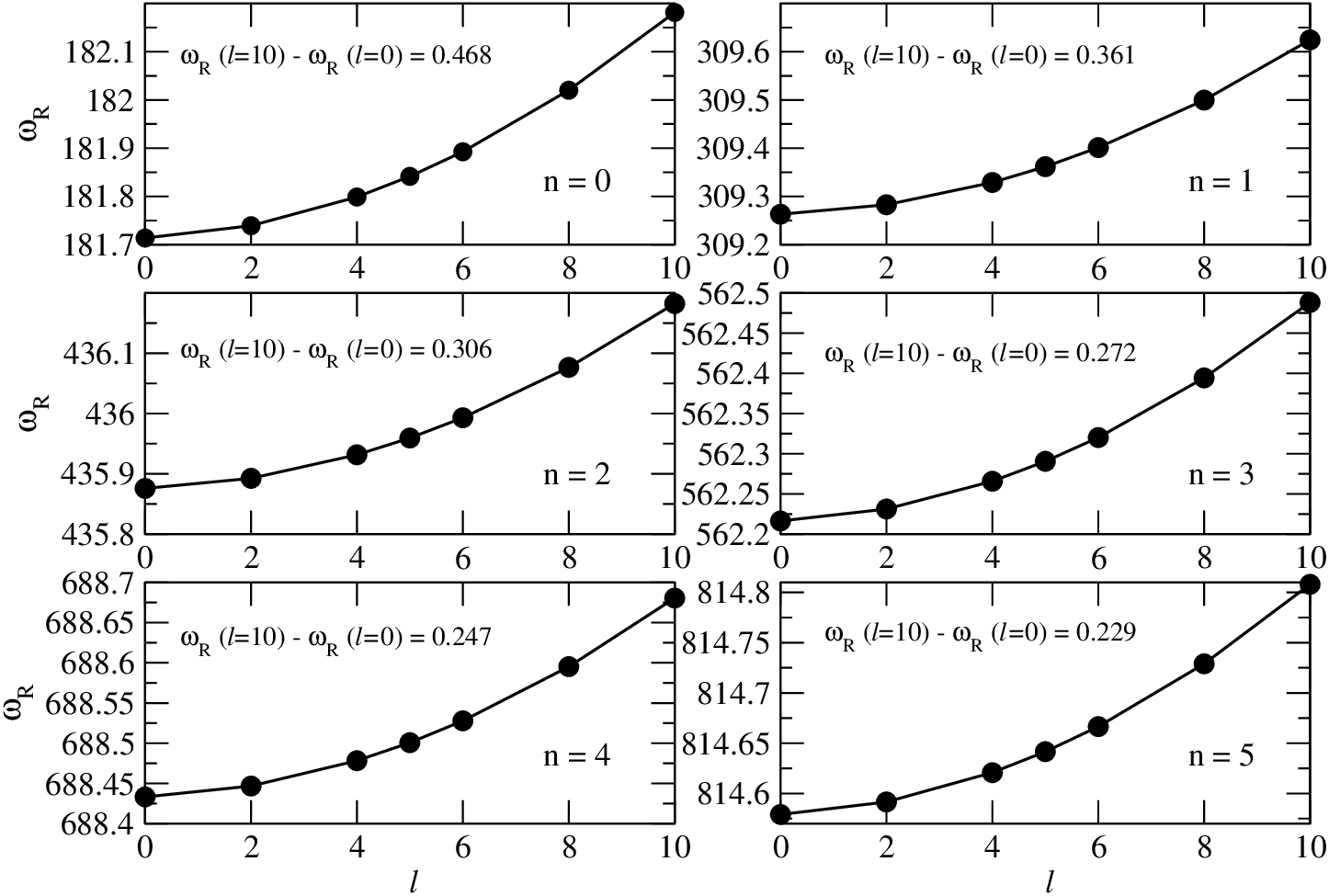}} 
\caption{Graphs of $\omega_{R} (OM)$  with
  $\ell$, for several values of $n$. The bullets indicate the
  numerical results and the lines indicate the quadratic fittings. For
  the curves, $r_{+}=100$, $R=1$ and $Q/Q_{max}=0.1$.}  
\label{fig11}    
\end{figure}

\begin{figure}
\resizebox{1\linewidth}{!}{\includegraphics*{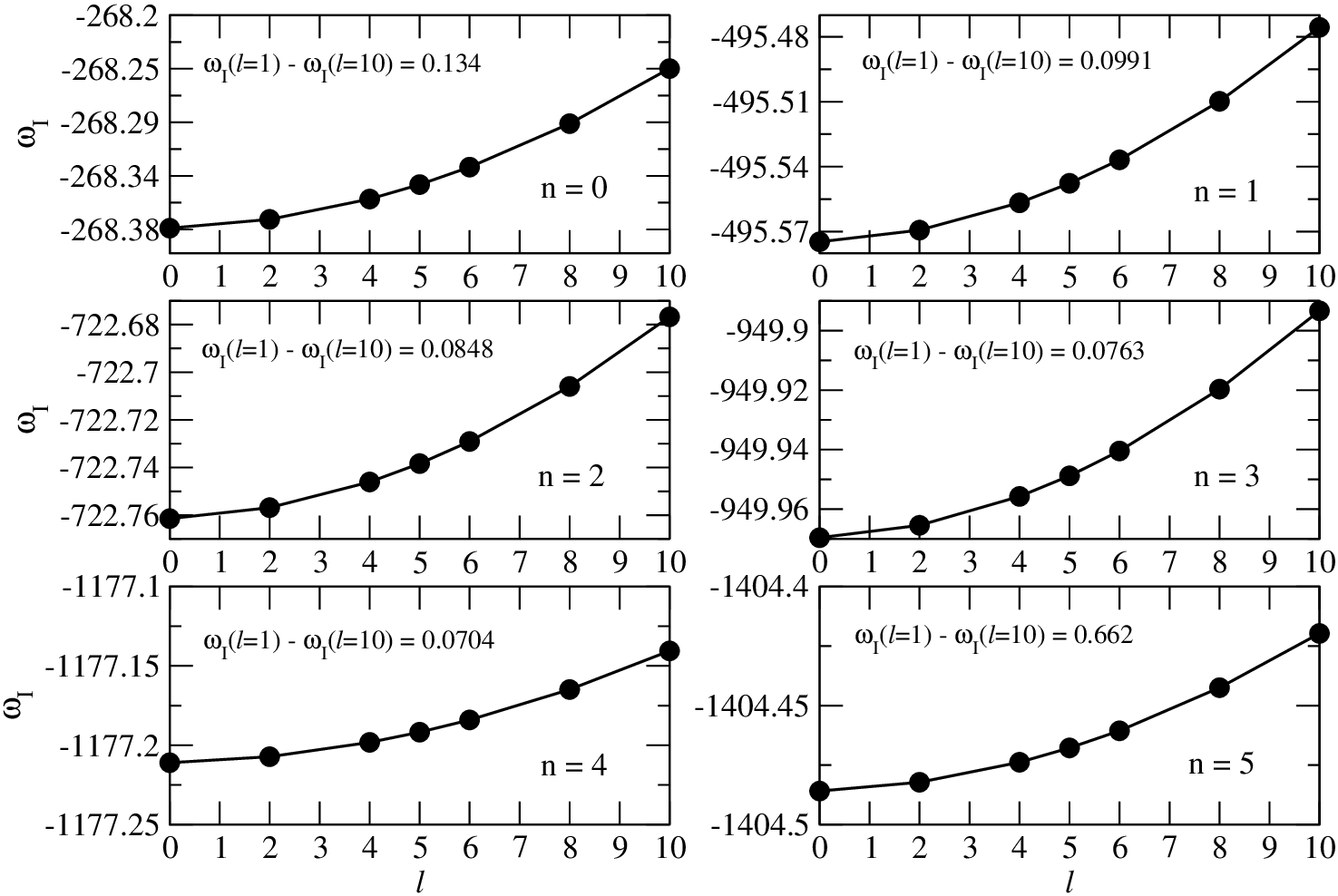}} 
\caption{Graphs of $\omega_{I} (OM)$  with
  $\ell$, for several values of $n$. The bullets indicate the
  numerical results and the lines indicate the quadratic fittings. For
  the curves, $r_{+}=100$, $R=1$ and $Q/Q_{max}=0.1$.}  
\label{fig12}    
\end{figure}

\begin{figure}
\resizebox{1\linewidth}{!}{\includegraphics*{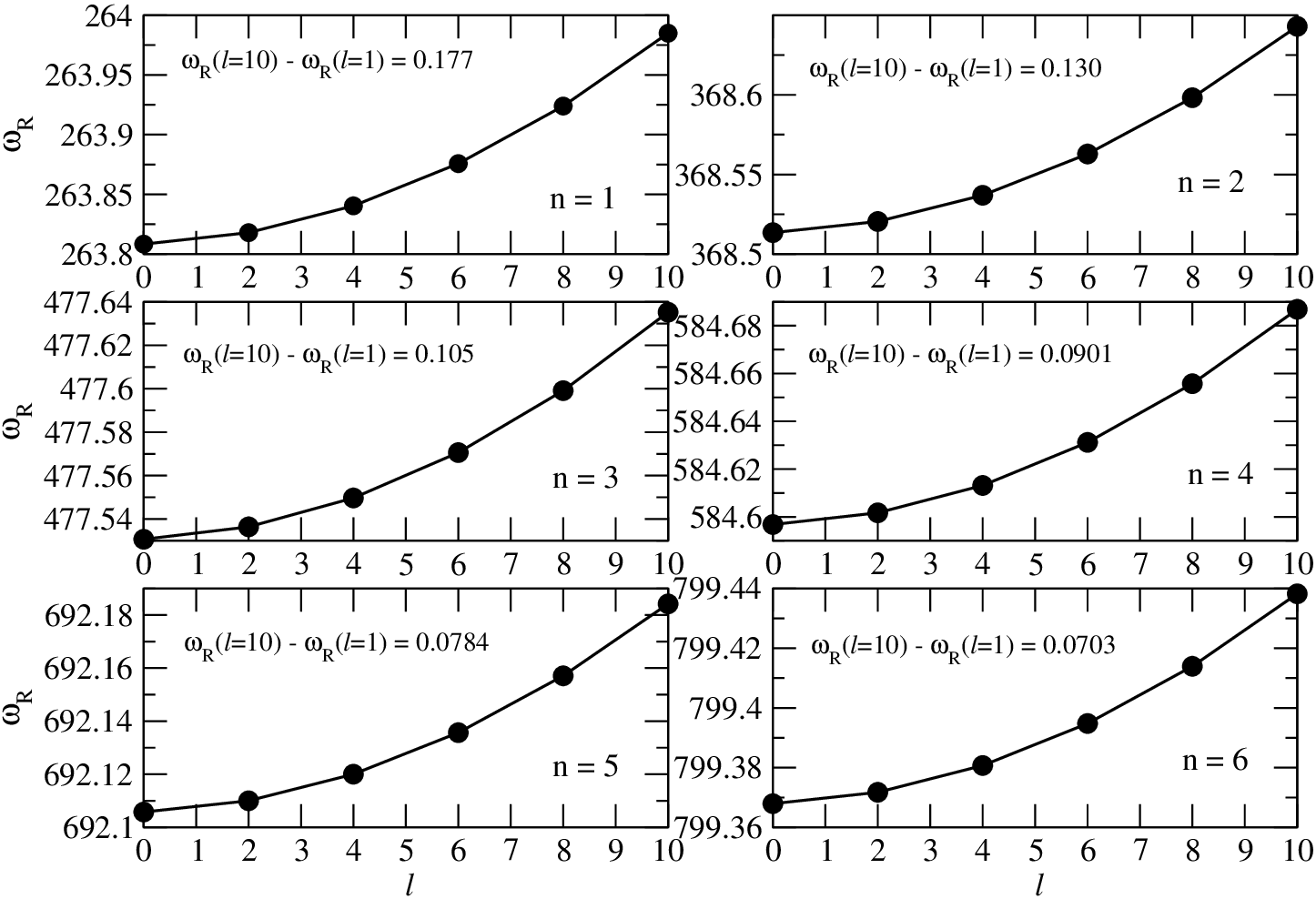}} 
\caption{Graphs of $\omega_{R} (OM)$  with
  $\ell$, for several values of $n$. The bullets indicate the
  numerical results and the lines indicate the quadratic fittings. For
  the curves, $r_{+}=100$, $R=1$ and $Q/Q_{max}=0.4$.}  
\label{lxReW_Qr0.4}    
\end{figure}

\begin{figure}
\resizebox{1\linewidth}{!}{\includegraphics*{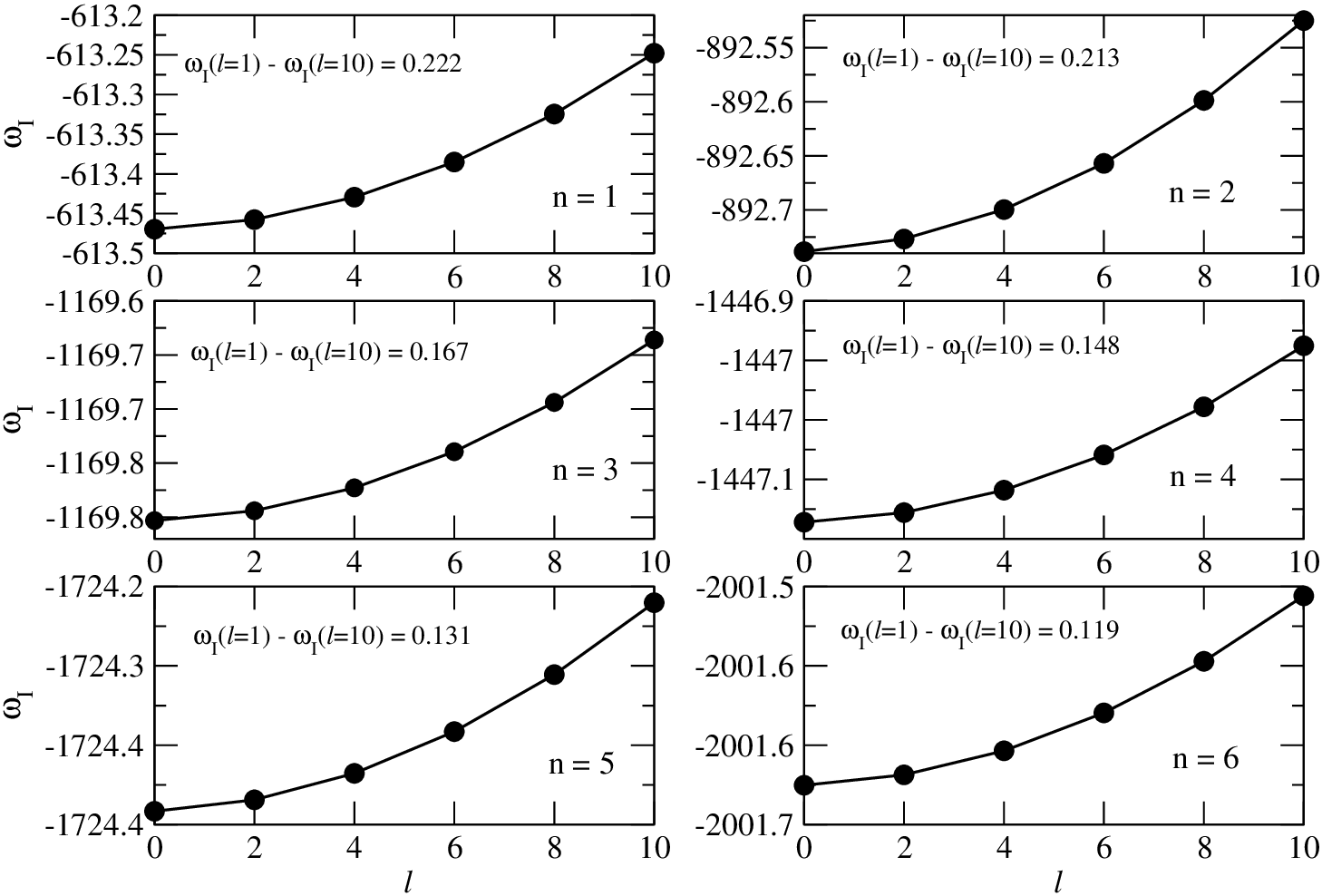}} 
\caption{Graphs of $\omega_{I} (OM)$  with
  $\ell$, for several values of $n$. The bullets indicate the
  numerical results and the lines indicate the quadratic fittings. For
  the curves, $r_{+}=100$, $R=1$ and $Q/Q_{max}=0.4$.}  
\label{fig14}    
\end{figure}

Our numerical computations show that although QNM frequencies depend
very weakly on $\ell$, the dependence is not trivial. As was found in
the lowest lying mode, $\omega_R$ increases and $\vert\omega_I\vert$
decreases with the increase of $\ell$. Besides, a
quadratic adjustment fits very well the dependence of the modes will
$\ell$, as seen in Fig.\ref{fig11}-\ref{fig14}.  
This property holds independently of the values of charge and the
overtone number we calculated. However we observed that $\omega_R$
increases and $\vert\omega_I\vert$ decreases slower with $\ell$ when
$n$ increases.

%%%%%%%%%%%%%%%%%%%%%%%%%%%%%%%%%%%%%%%%%%%%%%%%%%%%%%%%%%%%
%%%%%%%%%%%%%%%%%%%%%%%%%%%%%%%%%%%%%%%%%%%%%%%%%%%%%%%%%%%%

%%%%%%%%%%%%%%%%%%%%%%%%%%%%%%%%%%%%%%%%%%%%%%%%%%%%%%%%%%%%
%%%%%%%%%%%%%%%%%%%%%%%%%%%%%%%%%%%%%%%%%%%%%%%%%%%%%%%%%%%%

%%%%%%%%%%%%%%%%%%%%%%%%%%%%%%%%%%%%%%%%%%%%%%%%%%%%%%%%%%%%
%%%%%%%%%%%%%%%%%%%%%%%%%%%%%%%%%%%%%%%%%%%%%%%%%%%%%%%%%%%%

%%%%%%%%%%%%%%%%%%%%%%%%%%%%%%%%%%%%%%%%%%%%%%%%%%%%%%%%%%%%
%%%%%%%%%%%%%%%%%%%%%%%%%%%%%%%%%%%%%%%%%%%%%%%%%%%%%%%%%%%%

%%%%%%%%%%%%%%%%%%%%%%%%%%%%%%%%%%%%%%%%%%%%%%%%%%%%%%%%%%%%%%%%%%%%%%%%%%%%%%%
\section{Conclusions and discussions}
%%%%%%%%%%%%%%%%%%%%%%%%%%%%%%%%%%%%%%%%%%%%%%%%%%%%%%%%%%%%%%%%%%%%%%%%%%%%%%%

We have presented a comprehensive study of scalar perturbations of a
Reissner-Nordstr\"{o}m anti-de Sitter spacetime, and computed its
quasinormal modes. We observed the presence of two families of
solutions, both physically relevant. For $Q$ smaller than a critical
value, the asymptotic decay is dominated by  oscillating modes, while
for larger values of $Q$, the decay is dominated by  purely
exponential modes. Similar situation was observed in the
Reissner-Nordstr\"{o}m de Sitter (RNdS) geometry, where  $\Lambda$ was
the critical parameter [4]. 

For the lowest lying QNMs, our results verify (or sometimes disprove)
and extend many results obtained in
\cite{Horowitz-00,Wang-00,Wang-01,Cardoso,Berti-03}. The real part of the QN
frequency does have a local minimum at $Q/Q_{max}=0.366$, however when
$Q>0.3895Q_{max}$, the NOM starts to dominate. 
For this class of quasinormal modes, the real part of frequency vanishes. 
When selecting the mode class (OM or NOM) which dominates the decay,
we found that the imaginary part of the QN frequency
$\vert\omega_I\vert$ has  maxima at the same values of charge where
$\omega_R$ has minima. Before these maxima, $\vert\omega_I\vert$
increases monotonically with charge (OM solution), while after that
$\vert\omega_I\vert$ keeps on decreasing until the black hole extreme
limit (NOM solution). We observed that under the scalar perturbation
the imaginary part of the frequency tends to zero, as previously
conjectured in the electromagnetic and axial gravitational
perturbations \cite{Berti-03}. In the extreme limit, the asymptotic
field decay is dominated by a power-law tail, which suggests that this
geometry is stable to scalar perturbations. A similar field behavior
was observed in the RNdS case \cite{deSitter_2} in the approach to
$\Lambda=0$ limit.     

We have done an extensive search for higher overtones $n$ of the QNMs
of RNAdS black hole corresponding to scalar perturbations. We have
observed large variations of QN frequencies with the overtone number
and black hole charge. Both real and imaginary parts of QN frequencies
are evenly spaced for high overtone number $n$ at fixed $Q$. The
spacings between frequencies varies with the black hole charge. The
nontrivial quadratic dependence of QN frequencies on angular index $\ell$ has
also been addressed.

%%%%%%%%%%%%%%%%%%%%%%%%%%%%%%%%%%%%%%%%%%%%%%%%%%%%%%%%%%%%%%%%%%%%%%%%%%%%%
\begin{acknowledgments}
%%%%%%%%%%%%%%%%%%%%%%%%%%%%%%%%%%%%%%%%%%%%%%%%%%%%%%%%%%%%%%%%%%%%%%%%%%%%%

B. Wang's work was partially supported by  NNSF of China, Ministry of
Education of China and Shanghai Science and Technology
Commission. The work of Chi-Yong Lin was supported in part by the
National Science Council under the Grant
NSC-93-2112-M-259-011. C. Molina's work  was partially supported by
FAPESP, Brazil.

\end{acknowledgments}

%%%%%%%%%%%%%%%%%%%%%%%%%%%%%%%%%%%%%%%%%%%%%%%%%%%%%%%%%%%%%%%%%%%%%%%%%%%%%%%

\end{document}